# Physical improvements to a mesoscopic cellular automaton model for three-dimensional snow crystal growth

James G. Kelly and Everett C. Boyer
Centre College, Danville, KY 40422

**Abstract:** We motivate and derive the dynamical rules for a computationally feasible three-dimensional cellular automaton model of snow crystal growth. The model improves upon points of weak physical connections identified in other similar models which have produced morphological features observed in many snow crystal photographs. A systematic survey of the morphologies resulting from our model illustrates the degree to which these features persist in our results, and the trends that appear as model parameters are varied.

## 1. Introduction

Snow crystals exhibit a wide variety of faceted, dendritic, plate-like and needle-like morphologies [1, 2, 3, 4], and it is natural to speculate on the microscopic and dynamic origins of these forms. The variation in morphology with growth environment was systematically investigated by Nakaya [5], but progress in understanding the dynamic origins of crystal features was slow until computers became widely available for numerical simulation. Early hints at the possible origins of the morphologies came from computer experiments with diffusion-limited aggregation [6, 7, 8], which showed that the restriction of growth by vapor diffusion could result in complex, dendritic growth patterns that are qualitatively characteristic of some snow crystals. Later the front tracking and phase field approaches were developed [9, 10], which more directly treat the underlying diffusion with moving crystal boundary. Some recent results have shown promise [11], but achieving even qualitative agreement with observed morphologies has been difficult in cases where faceting is important [12].

Recent local cellular automaton (LCA) based models of snow crystal growth [13] have displayed an impressive ability to generate morphological features observed in many snow crystal photographs. In particular, the three-dimensional Gravner-Griffeath model [14] has provided a spectacular demonstration of the expressive power of cellular automata, generating virtual snow crystals with features that other modeling techniques have struggled to produce. Such qualitative agreement provides hope for future insight into the possible dynamic origins of some crystal surface features.

Computer models of snow crystal growth face two major challenges. First, the essential physics of molecular attachment at the crystal surface remains largely unknown [12, 15]. Second, even if the molecular-scale physics were known, the problem of modeling crystal growth to millimeter scales on a computer appears unfeasible because of the wide scale separation. The Gravner-Griffeath model avoids the above computational feasibility issue by defining the dynamics at a mesoscopic (roughly micron) scale intermediate between the nanoscale and the millimeter scale, so that millimeter-scale growth can be realized. It addresses the problem of poorly understood attachment physics in two ways. First, the model is strongly constrained by a wisely-chosen modeling grid that shares the symmetries of the underlying ice crystal, and which greatly influences the crystal growth process. Second, the dynamical rules that are introduced on this grid include many free parameters which are only loosely constrained and which admit only rough physical interpretations. The sizeable parameter space is then searched for promising morphologies with the hope that the unspecified physics is somehow captured within some region of the free parameters.

As suggestive as the resulting crystal shapes are, it would be premature to take them too literally in light of the relatively obscure connection of the model with crystal surface processes. An attempt to connect LCA models more directly to the physical attachment processes that are known was made Libbrecht in [16], but a fully three-dimensional simulation was not attempted. In this paper we build on some of the methods suggested by Libbrecht's model, identifying features of the Gravner-Griffeath model that can be altered to make the underlying model more physically realistic. We then develop a computationally feasible mesoscale model, and the resulting morphologies are systematically explored.

Section 2 briefly describes commonly-used ingredients in LCA models of snow crystal growth, including those implemented in Gravner-Griffeath and Libbrecht models. We also summarize specific aspects of the Gravner-Griffeath method on which we will attempt to improve. Section 3 describes our model in detail, including the motivation for our somewhat different dynamical approach involving alternating vapor relaxation and surface growth dynamics. Section 4 investigates the morphologies resulting from our model as the condensation coefficients and ambient vapor supersaturation are varied. Section 5 summarizes our conclusions.

## 2. Ingredients for cellular automaton models of crystal growth

We begin with a sketch of the commonly-used ingredients for an LCA model of snow crystal growth: the properties of the LCA grid of cells, the specification of the state of each cell and its interpretation, and the dynamics that specify how the state of each cell changes as the simulation progresses. Special attention is paid to the specific Gravner-Griffeath [14] and Libbrecht [16] approaches.

### 2.1. Choice of cellular automaton grid

The grid underlying the LCA partitions three-dimensional space into cells, each of which may contain ice crystal, supersaturated air, or both as in the case of a crystal boundary region. It is wise to choose a grid geometry that respects the symmetries of the underlying crystal in order to more easily accommodate surface-geometry-dependent growth speeds and facetization. For ice $I_h$ this suggests a grid of hexagonal prisms, for which a handful of cells is shown in Fig. 1*a*. One could implement dynamics on such a grid without explicitly specifying the inter-cell spacings $\Delta z$ and $\Delta x$, but specifying definite values for both allows a more direct physical interpretation of the resulting dynamics. The Gravner-Griffeath and Libbrecht models both take $\Delta z = \Delta x$. The Gravner-Griffeath model implements diffusion and attachment dynamics that make no explicit reference to $\Delta x$, and it is left unspecified in simulations. The Libbrecht model relates diffusion and attachment processes to Δ*x*, and it becomes a free parameter to be explored in simulations.

An LCA also defines a small neighborhood of adjacent cells for each cell and the dynamics of a cell is specified only in terms of cell states in this neighborhood. A very common definition for the neighborhood of a cell *C* consists of the cells 1-8, shown in Fig. 1*b*, in which the central cell *C* is obscured at the center. Both Gravner-Griffeath and Libbrecht models make this choice, and while other neighbor structures may offer more flexible growth dynamics, this definition seems to be both reasonable and minimal. In what follows we refer to cells 1-6 as the horizontal neighbors of *C* and cells 7-8 as the vertical neighbors.

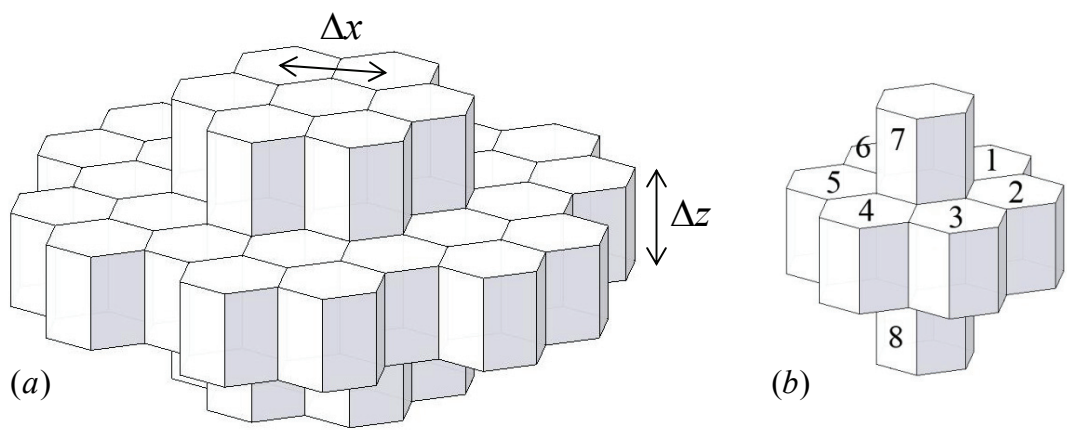

**Fig. 1.** (*a*) A grid of hexagonal prisms showing the horizontal cell spacing Δ*x* and the cell height Δ*z*; (*b*) The various neighbors 1-8 of a cell *C*, which is obscured at the center, are shown.

## 2.2. Cell types

The state of the LCA at a certain time must allow for an easy interpretation of both the spatial extent of the crystal and the distribution of water vapor outside of the crystal. To accomplish this, the Gravner-Griffeath and Libbrecht models have three different cell types. Those cells that are completely filled with crystal are referred to below as crystal cells. Those that contain only vapor-supersaturated air are referred to below as vapor cells. Those that are partially filled with both crystal and vapor and which contain part of the crystal surface are referred to below as boundary cells. We will see below that the vapor cells can convert to boundary cells, which can later convert to crystal cells, so that one important part of the state of the system is the knowledge of which cells are of which type. The crystal shape at any time is roughly the volume of space occupied by crystal cells.

## 2.3. Dynamic rules in the time-stepping method

Here we briefly describe the dynamical approaches used in the Gravner-Griffeath and Libbrecht models. The adjective "time-stepping" refers to idea that the state of each cell is updated in a way that simulates the exchange of vapor between neighboring cells, the uptake of vapor at the crystal surface, and the growth of the crystal that would happen during a small step forward in time. The dynamics of the crystal, vapor, and boundary cells types are necessarily quite different, so we describe each separately. The Gravner-Griffeath and Libbrecht approaches share many common features. The most crucial difference lies in how boundary cell dynamics are implemented. Below we summarize the common dynamical features of both models while pointing out the main differences and referring the reader to the individual papers for details. Our own dynamical approach, which differs somewhat from time-stepping, is described in section 3.

**Crystal cell dynamics**

Both the Gravner-Griffeath and Libbrecht approaches model only a period of continual growth without large amounts of melting or sublimation. Accordingly it is assumed that the ice-filled crystal cells do not evolve further in time, and the knowledge that a certain cell is a crystal cell is a complete description of its state.

**Vapor cell dynamics**

The state of a vapor cell consists of the water vapor density in the region (or the equivalent supersaturation, assumed positive). Vapor cells exchange water vapor with neighboring cells in accordance with a finite-difference approximation to the diffusion equation. A choice for the time step $\Delta t$ (along with a choice $\Delta x$ for the cell size) yields a rule for the new vapor density in a cell in terms of the current vapor densities in its neighborhood and the diffusion constant. The derivation of this approximation on a hexagonal prism grid is presented in [16], wherein careful consideration of numerical stability and fidelity of the entire algorithm informs the choice of $\Delta t$. The vapor exchange rule in the Gravner-Griffeath model makes no reference to a physical time step or cell size but is consistent with a fast but numerically stable finite-difference approach to vapor diffusion.

**Boundary cell dynamics**

Boundary cells contain both supersaturated air and crystal, so the state of a boundary cell consists of the local vapor density/supersaturation along with the mass of water in the cell in the crystal or quasiliquid state. Boundary cells exchange vapor with all non-crystal neighbor cells in accordance with the same rule used by vapor cells, with reflecting boundary conditions used when a neighbor site is a crystal cell. Boundary cells must also allow for conversions between water vapor and crystal/quasiliquid mass. In Libbrecht's model the Hertz-Knudson formula determines the rate with which the crystal gains mass, while the local vapor supersaturation is correspondingly reduced at a rate consistent with continuity. The Hertz-Knudson conversion rate depends explicitly on the local vapor supersaturation and a condensation coefficient $\alpha$ for the cell which parameterizes how readily vapor is incorporated into the crystal. In the Gravner-Griffeath the uptake of vapor into the crystal/quasiliquid mass is implemented by a linear model with coefficients that depend on the local crystal geometry, with no explicit mass continuity. There is also a similar linear model of melting which applies only to boundary cells.

**Cell type conversion**

When a boundary cell becomes filled with crystal the model must have a rule for converting the cell type to a crystal cell, and then converting adjacent vapor cells to boundary cells. In the Libbrecht model the algorithm checks at each time step to see of the accumulated crystal mass corresponds to a volume that would exceed the hexagonal prism volume. If it does the cell is converted to a crystal cell and any adjacent vapor cells are converted to boundary cells with no accumulated crystal mass. In the Gravner-Griffeath model the algorithm checks at every time step to see if the current quasiliquid mass in the cell exceeds a certain free model parameter $\beta$ that depends on the local crystal geometry. If it does the cell is converted to crystal type and newly adjacent vapor cells are converted to boundary cell type.

**Initial conditions and far boundary conditions**

The simulations generally begin with a small cluster of filled crystal cells and with the adjacent boundary cells containing no crystal/quasiliquid mass. The boundary condition far from the crystal is chosen to model a constant ambient supersaturation growth environment. To achieve this condition the vapor supersaturation can be held at a constant value $\sigma_\infty$ on the outer grid boundary, which is placed far enough away from the crystal so that the crystal growth is not sensitive to the exact boundary shape or location. The simulation is stopped before the crystal surface has growth close enough to the outer grid boundary to artificially speed growth. The Gravner-Griffeath model instead uses a wrapped outer boundary condition and checks to make sure that the vapor density in the far cells does not fall much below its starting level for the duration of the simulation. The initial supersaturation in all non-crystal cells is set at $\sigma_\infty$.

**2.4. Physical deficiencies of cellular automaton models of crystal growth**

Of the two models considered above, the Gravner-Griffeath model has a weaker connection with physics but nonetheless produces a very suggestive set of morphological features that compare qualitatively well with photographs. We now list some of the potentially problematic features of this model which may cause one to wonder if the interesting morphological features persist if one alters the model in an attempt to increase fidelity.

1. The hexagonal prism cell size is not precisely related to the growth and diffusion processes. Almost any kind of physical strengthening of the model would seem to require that the grid scale be defined.

2. The supersaturation is drained to nearly zero in crystal boundary cells at every time step. We imagine that instead the supersaturation should remain fairly constant at the crystal surface, and it should change only by a small percent in any time step. We worry that this kind of overstepping in time may cause the average supersaturation level in the boundary cells to be artificially low, giving undue growth advantage to crystal parts that protrude slightly and thereby encouraging more dendritic growth or surface structure than is justified.

3. The speed of the crystal surface growth is loosely physically motivated. Libbrecht suggests the use of the Hertz-Knudson formula [16, 17] as a reasonable starting place for the dependence of growth rate on supersaturation level.

4. There is no continuity relation between mass gained by the crystal and the reduction in supersaturation. This may cause the crystal to grow at a rate that is too fast compared to the vapor transport rate. From a simulation point of view this is nice, because one the cells fill at a reasonable rate and somewhat large final crystal sizes are feasible. In [16] a justification for a careful decoupling between vapor transport rate and crystal growth rate is developed so that simulation times become reasonable while avoiding physically questionable overdriven crystal growth rates.

5. It is not clear that the separate quasiliquid and crystal masses in a boundary cell are crucial for the dynamics. Our suspicion is that a single mass parameter would suffice, especially in the absence of the problems in items 2 and 4.

6. An artificially high initial condition for the vapor density near the crystal may fuel initial growth, which can affect long-term morphology. Our own simulations have shown that in order to model an ambient supersaturation "at infinity" one must place a constant supersaturation outer boundary quite far from the crystal. A model involving a wrapped outer boundary condition and constant supersaturation initial condition that is run until the supersaturation at the far boundary falls by a small fraction may be feeding vapor to the crystal surface at too great a rate for much of the simulation time, with a potentially large effect on the final morphology.

7. There are many free parameters in the model. Ideally this number would be somewhat reduced in a more physically motivated model.

In fact Libbrecht's model addresses most of these issues, but leaves open the question of whether the resulting interesting morphological features in three-dimensional growth are altered. We now turn to the development of a model in which we attempt to address these issues while retaining feasible simulation times for three-dimensional crystals that are large enough to develop characteristic features seen in photographs.

## 3. A physically motivated cellular automaton approach to crystal growth

We now develop a three-dimensional LCA crystal growth model that attempts to address many of the issues raised in section 2. We begin with a discussion of the rules for vapor diffusion, crystal surface growth, and vapor removal at the crystal surface in the context of the time-stepping method. Beginning in section 3.4 we derive the rules for the alternating growth-relaxation dynamical method.

### 3.1. The finite difference rule for vapor diffusion on a hexagonal prism grid

A vapor diffusion rule can be obtained from the diffusion equation by making discrete approximations to the spatial and time derivatives. There are many ways to do this; we use the dynamical rule derived in [16] which is stated in terms of the vapor supersaturation $\sigma = (c - c_{sat})/c_{sat}$, where $c$ is the water molecule number density in a cell and $c_{sat}$ is the number density at saturation. The rule is

$$\sigma_{ctr}(t + \Delta t) \approx \tfrac{2}{3} \Delta \tau \sum_{i=1}^{6} \sigma_i(t) + \Delta \tau \sum_{j=1}^{2} \sigma_j(t) + (1 - 6\Delta \tau)\sigma_{ctr}(t), \qquad (1)$$

where $\Delta t$ is the discrete time step, $\Delta \tau = D\Delta t / (\Delta x)^2$ is a dimensionless time step defined in terms of the diffusion constant $D$ and the grid cell separation $\Delta x = \Delta z$, $\sigma_{ctr}$ is the vapor supersaturation in the cell being updated, $\sigma_i$ are the supersaturation values in the horizontal neighbor cells, and the $\sigma_j$ are those in the vertical neighbor cells. We use reflecting boundary conditions at the crystal surface: if one or more of a vapor-containing cell's neighbors are crystal cells we substitute $\sigma_{ctr}$ for each $\sigma_i$ or $\sigma_j$ where this occurs.

## 3.2. The rule for crystal surface growth

We wish to find the amount of ice volume accumulated by a cell in a time step $\Delta t$ in the presence of local supersaturation $\sigma$. We begin with the Hertz-Knudson formula [16, 17] for the normal growth speed of the ice surface,

$$v_n = \alpha \frac{c_{sat}}{c_{solid}} \sqrt{\frac{kT}{2\pi m}} \sigma = \alpha v_{kin} \sigma, \qquad (2)$$

where $c_{solid}$ is the water molecule number density in ice, $kT$ is Boltzmann's constant times temperature, $m$ is the water molecule mass, and $\alpha$ is a dimensionless condensation coefficient which parameterizes how readily a water molecule that contacts the crystal is incorporated into the crystal lattice. The formula assumes that $0 \leq \alpha \leq 1$ where $\alpha = 1$ means that every water molecule that encounters the crystal is immediately incorporated into the lattice, rather than migrating on the crystal surface or being subsequently removed by thermal motion.

Very roughly one can imagine that the ice surface progresses forward at speed $v_n$ within a grid cell. We wish to know the rate at which the cell accumulates ice volume since this determines how quickly the cell fills with ice, and because vapor must be removed from the air at a corresponding rate. Unfortunately there are many ways in which the ice might fill a cell. The most reasonable assumption seems to be that a flat sheet of ice progresses across the cell in a certain direction at constant speed $v_n$, but this would give a nonconstant ice volume accumulation rate in the cell as the ice sheet moves from a corner of the cell to a central region with a larger cross section. We believe that any sensible approach should have a constant ice volume accumulation rate which is correct on average, but even this "correct" average rate depends on the assumed direction of ice sheet movement for any non-spherical cell shape. Fortunately our assumed $\Delta x = \Delta z$ hexagonal prism cell shape is not too far from being spherically symmetric, so the range of variation of average mass accumulation rates is small. We make a particular choice in the derivation below, and we keep in mind that a slightly different choice would result in the same dynamics at a slightly different $\alpha$ value.

If we want the crystal volume to change at the correct average rate then the change in crystal volume within the cell during time step $\Delta t$ must be

$$\Delta V_{avg} = \frac{V_{cell}}{T} \Delta t, \qquad (3)$$

where $V_{cell} = \frac{\sqrt{3}}{2}(\Delta x)^3$ is the grid cell volume and $T$ is the time it takes to fill the cell. We make the particular choice of a single ice plane progressing across the cell perpendicular to a prism facet direction at speed $v_n = \alpha v_{kin} \sigma$. The distance between adjacent facet cell planes in this direction is $\frac{\sqrt{3}}{2}\Delta x$ so that $T = \frac{\sqrt{3}}{2}\Delta x / (\alpha v_{kin} \sigma)$. With these substitutions in equation ( 3 ) we obtain

$$\Delta V_{avg} = \alpha \sigma v_{kin} (\Delta x)^2 \Delta t. \quad (4)$$

### 3.3. The rule for vapor removal at the crystal surface

Continuity requires that the accumulation of ice mass at the crystal surface given by equation ( **4** ) be balanced by a corresponding removal of vapor from the air. If the ice volume in a cell changes during a time step by $\Delta V$, then the number of water molecules in the crystal increases by

$$\Delta N = c_{solid} \Delta V, \quad (5)$$

where $c_{solid}$ is the number density of ice. In the air the number of molecules changes by $-\Delta N$, and the change in the supersaturation $\Delta \sigma$ in the same cell is

$$\begin{aligned}\Delta \sigma &= \frac{c_{t+\Delta t} - c_{sat}}{c_{sat}} - \frac{c_t - c_{sat}}{c_{sat}} \\ &= \frac{(N_{air, t+\Delta t}/V_{cell}) - (N_{air, t}/V_{cell})}{c_{sat}} \\ &= \frac{-\Delta N}{c_{sat} V_{cell}} = \frac{-c_{solid} \Delta V}{c_{sat} V_{cell}},\end{aligned} \quad (6)$$

where $c_{sat}$ is the number density of water vapor at saturation just above an infinite plane of ice, $c_t$ and $c_{t+\Delta t}$ are the number densities of water molecules in the crystal boundary cell before and after the removal of the molecules that join the crystal, and $V_{cell}$ is the volume of the grid cell. Combining with the volume accumulation rule ( 4 ) above we obtain

$$\begin{aligned}\Delta \sigma &= \frac{-c_{solid} \alpha \sigma v_{kin} (\Delta x)^2 \Delta t}{c_{sat} \frac{\sqrt{3}}{2}(\Delta x)^3} \\ &= -\alpha \sigma \frac{2}{\sqrt{3}} \sqrt{\frac{kT}{2\pi m}} \frac{\Delta t}{\Delta x} \\ &= -\alpha \frac{2\Delta x}{\sqrt{3}D} \sqrt{\frac{kT}{2\pi m}} \sigma \Delta \tau \\ &\equiv -K \sigma \Delta \tau,\end{aligned} \quad (7)$$

where the last equality defines the draining constant $K$ for a boundary cell. For later convenience we define $K$ to be zero in a vapor cell since it contains no supersaturation sink.

### 3.4. Motivation for decoupling the vapor diffusion and crystal growth processes

Above we have presented the three fundamental modeling ingredients for stepping the cellular automaton forward in time. The most straightforward way to run a simulation is to choose an appropriately small time step $\Delta t$ and implement these rules in each vapor or boundary cell at each time step, holding the supersaturation constant at the far boundary and making cell conversions from vapor to boundary type or from boundary to crystal type as cell fillings occur. Since the goal of the simulation is to grow the crystal out for a certain length of time one may wish to choose a larger time step to make the simulation proceed faster, but too large of a step may also introduce numerical errors or instabilities into the simulation. We now consider the scale of the time step required to have a stable simulation of reasonable fidelity.

The two relevant time scales for the ice crystal growth problem are discussed in [16]. The time scale for diffusion to adjust the vapor concentration in the vicinity of the crystal is $\tau_{\text{diffusion}} \approx R^2 / D$, where $R$ is the approximate crystal radius. The time scale for crystal growth is $\tau_{\text{growth}} \approx 2R / v_n$, where $v_n$ is a typical normal growth velocity. The ratio of these scales is known as the Peclet number, $p = Rv_n / 2D$, and in [16] it is noted that for ice crystal growth in the atmosphere this is very small, typically $p \leq 10^{-5}$. This means that the crystal geometry changes very slowly while the vapor adjusts via diffusion on a much shorter times scale. For all practical purposes the supersaturation field is maintained in a state of equilibrium with respect to the crystal surface and far grid boundary conditions.

The significant difference in the time scales of the two processes creates a computational problem for the time stepping simulation approach outlined above. In order for the finite difference diffusion dynamics to be stable and of reasonable fidelity the time scale for the diffusion time step must be small compared to $\tau_{\text{diffusion}}$, but such a time step would make the growth proceed as a prohibitively slow rate. In order to allow for reasonable simulation times, the growth algorithm presented in [16] speeds up the ice volume accumulation rate by a factor of $\Lambda$ relative to the vapor removal rate to a point where the Peclet number is much larger but still less than 1.

A second consideration also limits the size of the time step. The draining rule ( 7 ) indicates that for a large enough time step the supersaturation may be drained to negative values. We wish to avoid this possibility because it may cause instabilities in the simulation. Moreover, for real crystal growth the supersaturation ought to stay relatively constant at the crystal surface rather than fluctuating greatly within one time step. A more conservative method would require that only a small fraction $\Delta\sigma/\sigma$ of the supersaturation is removed due to vapor removal during a time step.

In our experience the simulation times involved in using the time stepping approach with growth acceleration $\Lambda$ to meet this criterion are still prohibitively large, so we take a different approach. The small Peclet number indicates that the supersaturation field is effective in equilibrium and so does not change with time. In this limit the diffusion equation reduces to Laplace's equation. Our approach, to which we refer here as the alternating growth-relaxation method, starts with a given crystal geometry and solves Laplace's equation via relaxation with the correct boundary conditions at the crystal surface and on the far grid boundary. Then, knowing the supersaturation level in all cells on the crystal boundary, we grow the crystal in each of the boundary cells until one of the cells is filled. This can be done in a single step. After turning on the new adjacent boundary cells, we again find the Laplace equation solution for the supersaturation with the slightly changed crystal geometry, and then we grow again until boundary cell fills. In this way we alternate between growth and relaxation steps, with all of the real computational time spent in the relaxation phase. As the crystal shape has only changed slightly in a very small region for each new relaxation step, the previous supersaturation field provides an excellent initial guess for the new iterative relaxation process.

### 3.5. The rule for relaxation of the supersaturation field: *α* independent of *σ* case

One can think of the dynamics that determine the vapor supersaturation as composed of two steps. First diffuse according to ( 1 ) using reflecting boundary conditions, then remove vapor from crystal boundary cells according to ( 7 ):

$$\sigma' = \tfrac{2}{3}\Delta\tau \sum_{i=1}^{6} \sigma_i + \Delta\tau \sum_{j=1}^{2} \sigma_j + (1-6\Delta\tau)\sigma$$

$$\sigma'' = \sigma' - K\sigma'\Delta\tau.$$

(8)

The second step is trivial in vapor cells where $K = 0$, but is crucial for boundary cells. These can be combined into a single step:

$$\sigma'' = \left( \tfrac{2}{3}\Delta\tau \sum_{i=1}^{6} \sigma_i + \Delta\tau \sum_{j=1}^{2} \sigma_j + (1-6\Delta\tau)\sigma \right)(1-K\Delta\tau).$$

(9)

Because the diffusion is fast compared to the rate of crystal growth, *i.e.* the Peclet number is low, we can assume that the value of $\sigma$ in both pure vapor and crystal boundary cells will be in equilibrium and only change very slowly as the crystal grows. In equilibrium $\sigma$ will return to its same value after each time step so that $\sigma'' = \sigma$. If *α* is independent of *σ* so that $K$ is also independent of *σ* we may solve for the equilibrium value to obtain

$$\sigma = \frac{\left(\tfrac{2}{3}\Delta\tau \sum_{i=1}^{6} \sigma_i + \Delta\tau \sum_{j=1}^{2} \sigma_j\right)(1-K\Delta\tau)}{(K+6)\Delta\tau - 6K(\Delta\tau)^2}.$$

(10)

We may now take the $\Delta\tau \to 0$ limit to obtain

$$\sigma = \frac{\frac{2}{3}\sum_{i=1}^{6}\sigma_i + \sum_{j=1}^{2}\sigma_j}{K+6}. \quad (11)$$

This is the relation between the supersaturation in a cell and those of its neighbors which should hold when the vapor is in equilibrium for the case where $\alpha$ is independent of $\sigma$. We can find the $\sigma$ values in the vapor and boundary cells that collectively satisfy this condition by using the method of relaxation, iteratively updating $\sigma$ values on all grid cells by repeatedly applying this equation. Boundary cells use the relevant draining constant $K$, while vapor cells use $K = 0$.

### 3.6. The rule for relaxation of the supersaturation field: $\alpha$ dependent on $\sigma$ case

In this case the equilibrium solution to ( 9 ) depends on the relationship between $\alpha$ and $\sigma$. We explore one simple kind of dependence here, based on discussions in [12]. When a facet is flat at the molecular level, individual adsorbed water molecules are only bound very weakly to the surface. Thermal motion tends to release them back into the vapor before they can be incorporated into the crystal. However, if the vapor density is large enough then the density of admolecules is such that islands of several admolecules form, which tend to stabilize each other and provide kink and step sites for enhanced binding of other vapor molecules. This process is known as nucleation-limited growth because net attachment rates are very small until a nucleation site is formed, and it is characterized by very suppressed condensation coefficients $\alpha$ at small $\sigma$ and $\alpha \to 1$ at large $\sigma$. The available growth data are fit very well by the formula [12, 16]

$$\alpha = \min(A(T,\sigma)e^{-\sigma/\sigma_0}, 1), \quad (12)$$

where the critical supersaturation $\sigma_0$ depends on the facet type, prism or basal. Apparently $A$ depends only weakly on $\sigma$ so that for growth at constant temperature we have

$$\alpha = \min(Ae^{-\sigma/\sigma_0}, 1) \quad (13)$$

for constants $A$ and $\sigma_0$.

Unfortunately after substitution of this relation into ( 9 ), the search for an equilibrium $\sigma$ involves solving a transcendental equation. To avoid the chore of generating numerical solutions, we approximate by the piecewise linear function

$$\alpha = \min(\sigma/(\sigma_0 \ln A), 1). \quad (14)$$

A comparison between the $\sigma$ dependence of ( 13 ) and ( 14 ) is shown in Fig. 2 for the values $A = 2$ and $\sigma_0 = 0.021$ corresponding to measured data for basal facet growth [16]. The approximation appears to capture the basic features of the variation, except at small supersaturation values where the approximation becomes rather poor. Our hope is that at low supersaturation the growth is sufficiently suppressed by our model relative to other crystal regions with higher surface supersaturation so that the overall morphology will not strongly be affected.

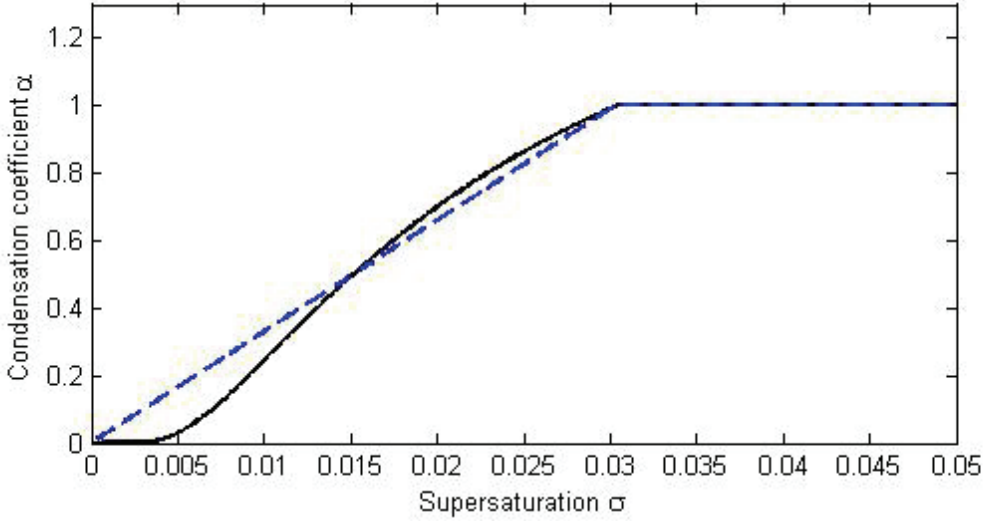

**Fig. 2. Measurement-based model of condensation coefficient variation with supersaturation for nucleation-limited basal facet growth (solid curve) and our piecewise-linear model (dashed curve).**

Substituting ( 14 ) into $K$ in ( 9 ) and solving for the equilibrium $\sigma$ value, we find that the equation is linear when $\alpha = 1$ and quadratic when $\alpha = \sigma/(\sigma_0 \ln A)$, allowing analytic solutions. Combining both cases together we obtain

$$\sigma = \begin{cases} \dfrac{-6 + \sqrt{36 + 4B\left(\frac{2}{3}\sum_{i=1}^{6}\sigma_i + \sum_{j=1}^{2}\sigma_j\right)}}{2B} & \text{if } \sigma < \sigma_0 / \ln A \\ \dfrac{\frac{2}{3}\sum_{i=1}^{6}\sigma_i + \sum_{j=1}^{2}\sigma_j}{K+6} & \text{if } \sigma \geq \sigma_0 / \ln A, \end{cases} \quad (15)$$

where

$$B = \frac{2\Delta x \ln A}{\sqrt{3} D \sigma_0} \sqrt{\frac{kT}{2\pi m}}. \quad (16)$$

We find relaxation solutions for the supersaturation field by iteratively applying equation ( 15 ), choosing the multipart formula based on the value of $\sigma$ in the center cell, until the field converges.

## 4. Simulation details

In this section we summarize our grid choice, state variables, dynamics, and physical constants used for the simulation results displayed in section 5.

### 4.1. Modeling grid

We use a grid of hexagonal prisms satisfying $\Delta z = \Delta x$, where $\Delta x$ is taken as a free parameter allowing the study of its effects on the crystal morphologies. For simplicity we would like to place the growing crystal at the center of the grid and hold the vapor supersaturation constant at the far boundary of the grid. Practically speaking this is not computationally feasible, as computational experiments have shown that the constant supersaturation boundary must be made extremely far from the crystal to avoid feeding anomalously fast crystal growth from a too-close boundary. To accommodate for this we use a two-part grid with a fine resolution region (the *fine grid*) close to the crystal, joined to a coarse resolution region (the *coarse grid*) that extends far from the crystal and has constant supersaturation at its far edge. The crystal itself is only allowed to grow within the fine grid, but vapor diffusion is modeled in both grid regions.

As another concession to computational feasibility we model only a perfectly symmetric snow crystal in a symmetric growth environment, so that the simulated fine grid includes only $1/24^{th}$ of the complete grid imagined above. The neighbors of cells that fall on one of the boundaries of this region that result from the division are determined by symmetry. The coarse and fine grids, with the vertical direction suppressed, are illustrated in Fig. 3. Note that only a 30 degree region of the plane is included in the simulation; the other locations can be computed by symmetry and are used for display purposes only. The lower left fine grid hexagon is centered on the center of symmetry of the crystal.

For storage in a C-language array we use a mapping similar to the one used in [16], where the first array dimension corresponds to the horizontal direction in the figure, the second array dimension corresponds to the direction diagonally upward to the right, and the third is out of the page. For comparison the figure illustrates a notional fine grid that could be stored in an array of size 17x9x…, where each coarse cell has three times the extent of a fine cell along each dimension. For our simulations we have most often used a fine grid array size of 141x72x22 cells, joined to a coarse grid array size of 61x32x62 cells, where the last dimension is the half-height of the basal growth direction and where a coarse cell has ten times the extent of a fine cell along each dimension. Such a grid is obviously not well-suited to thin, needle-like growth along the basal direction but the dimensions and relative sizes can be adjusted to accommodate other geometries.

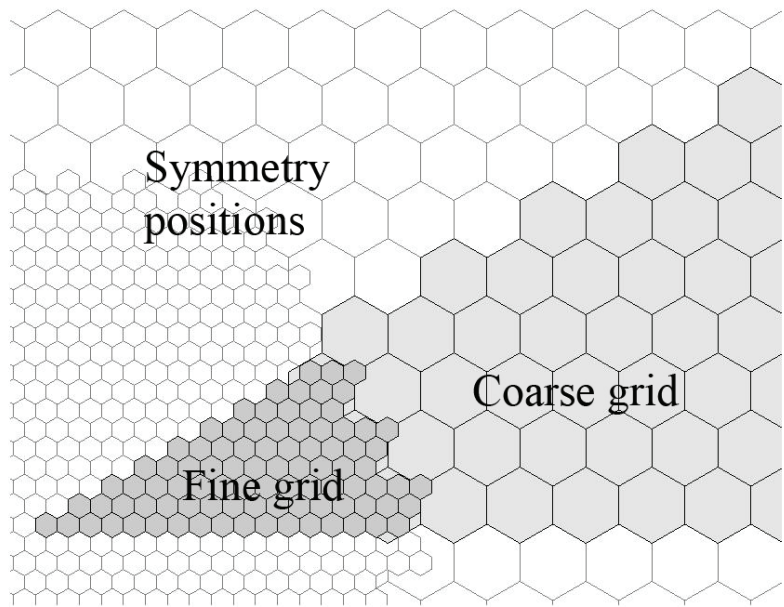

**Fig. 3. Notional diagram of coarse and fine grid (grey cells) used in the simulation, as well as the symmetry positions (white cells) used for visualization only.**

### 4.2. State variables and initialization

We have the crystal, vapor, and boundary cell types described in section 2. A fine grid cell can be any of these types, while coarse grid cells are only allowed to be vapor type. The state of the system consists of knowing the type of each cell, the water vapor supersaturation in each vapor cell, and the vapor supersaturation and length of the growing crystal in each boundary cell.

We initialize the fine grid with a small number of crystal cell sites and determine the neighboring boundary sites, which begin with crystal lengths of zero. For the simulation results in the next section we use a starting configuration of 21 crystal cells which is roughly in the shape of a hexagonal prism extending three cells in the vertical direction and with a diameter of three cells in the horizontal direction. This avoids the anomalous case of a single starting crystal cell with no filled neighbors, which computational experimentation has shown can lead to atypical initial growth and long-term morphologies. These experiments also indicate that there is some sensitive dependence on the initial crystal shape even for non-single-cell initial states. Our choice of a small hexagonal prism initial state is motivated by observation in [12] that most atmospheric snow crystals begin their growth as a small hexagonal prism, with branching instabilities occurring at larger sizes. While the cell size and out initial condition may forbid extremely early branching onset, we also do not expect our model to accurately capture the growth of snow crystals with features on the order of or smaller than the cell size.

### 4.3. Condensation coefficients

The Hertz-Knudson formula ( 1 ) plays a central role in both the crystal growth phase and the vapor relaxation phase (via the vapor removal boundary condition at the crystal surface) of our model. The normal growth rate depends on the condensation coefficient $\alpha$, which may depend on both the local supersaturation level via ( 14 ) and the local crystal geometry. Since a normal growth rate only makes sense for smooth surfaces and since ice crystals can be rough or kinked at the molecular scale, it probably makes little sense to ask what the *actual* dependence of $\alpha$ on the local geometry is; the Hertz-Knudson formula should only apply in the case of the flat local geometry at the center of a large, smooth facet.

A reinterpretation of the normal growth speed allows an extension the application of the Hertz-Knudson formula. Individual water molecules that adsorb onto the crystal are usually kicked off by thermal motion. If the molecule can create a stronger bond with the surface then it is more likely to stay on the surface long enough to be incorporated into the lattice, so that locations on the crystal that encourage strong bonding correspond to regions of greater growth rates. In general, the more crystal neighbors that an adsorbed molecule has in various spatial directions, the stronger the bond to the crystal surface and the faster the local growth rate. A molecule adsorbed onto the center of a smooth facet has a fairly weak bond, while one at an interior corner kink site has a much stronger bond and one at a protruding point makes a weaker bond. Although the cells of the LCA represent micron scale structures, not molecular-scale structures, if we see a kink site in the collection of LCA crystal cells we can conclude that a strong-bonding kink site must exist at the molecular level and will be quickly filled. The filling creates other adjacent strong-bonding sites, leading to an average fast rate of growth at larger scales which we can roughly model by assigning a large condensation coefficient $\alpha \approx 1$ to the kink-site cell. Similarly, a pointed tip location in the LCA crystal should be assigned a small $\alpha$, and so on.

A crude way to implement this bonding tendency is to count the number of filled crystal neighbor cells of a boundary cell and assign a relatively appropriate $\alpha$ value between 0 and 1. Since bonding along the horizontal and vertical directions can have different stabilizing effects, we count the number of horizontal and vertical filled neighbors separately. We denote the number of horizontal and vertical crystal neighbors of a boundary site via a subscript on the site's condensation coefficient, so that $\alpha_{HV}$ denotes a condensation coefficient for a boundary site with $H$ horizontal and $V$ vertical filled crystal cell neighbors. For example, if in Fig. 1*b* we imagine the displayed cells 1-8 as being the only crystal cells in the LCA then coefficient of the boundary cell above cell 7 would be denoted $\alpha_{01}$, while the one adjacent to cells 4 and 7 would be $\alpha_{11}$, and the one horizontally adjacent to cells 3 and 4 would be $\alpha_{20}$.

For simulations in which $\alpha$ is assumed to be independent of $\sigma$, the parameters $\alpha_{HV}$ are constants for the entire simulation. In this case we have chosen to explore a reduced two-dimensional parameter space of $\alpha_{01}$, (which controls the vertical growth rate) versus $\alpha_{10} = \alpha_{20} = \alpha_{11}$ (which control the horizontal growth rate), with all other condensation coefficients set equal to 1 due to large numbers of neighbors.

For simulations in which $\alpha$ is allowed to depend on $\sigma$ we limit the dependence to the parameters $\alpha_{01}$, $\alpha_{10}$, and $\alpha_{20}$, using a piecewise linear function modeling nucleation-limited growth as described in section 3.6. The kink in the curve illustrated in Fig. 2 will occur at different supersaturation levels for basal facet growth (*i.e.* vertical growth governed by $\alpha_{01}$) and for prism facet growth (*i.e.* horizontal growth governed by $\alpha_{10}$ and $\alpha_{20}$). For convenience we label the supersaturation level at which the kink occurs by $\sigma_{HV}$. For basal facet nucleation-limited growth we use the fitted values corresponding to measured growth rates given in section 3.6, which result in $\sigma_{01} \approx 0.03$. For prism facet nucleation-limited growth, for which [16] indicates that no good measurements exist, we set $\sigma_{10} = \sigma_{20}$ and treat this as a free parameter to be explored along with the ambient supersaturation $\sigma_\infty$. All of the other $\alpha_{HV}$ are held constant at 1 during the simulation.

### 4.4. Vapor relaxation and crystal growth

We relax the coarse and fine supersaturation field by repeatedly applying ( 11 ) or ( 15 ) across the coarse and fine grids one cell at a time. The field is normally considered to have converged when the equation sides differ by less than 0.005% in every coarse and fine cell. This convergence criterion was determined by computer experiment; by varying this number and growing out a crystal to full size we determined how large the number could be without a visually noticeable difference in the final crystal.

The normal growth rate of the crystal is computed from the Hertz-Knudson formula ( 2 ). Given the current crystal length and supersaturation value in each boundary cell we determine the smallest amount of time $\Delta t_{min}$ that would cause one of the boundary cells to fill. We then mark this cell as filled and turn on adjacent new boundary cells, and update the other boundary cell crystal lengths via ( 2 ). $\Delta t_{min}$ is added to the growth time accumulation variable.

### 4.5. Stopping criterion

The vapor relaxation and crystal growth steps are alternated until the crystal extent becomes too large for the fine grid. In the case of a fine grid array size of 141x72x22 cells we normally stop when the crystal reaches either a horizontal radius of 138 cells or a vertical half-height of 20 cells. The crystal cell occupation array, the supersaturation field, and the total growth time are then saved to disk for visualization.

### 4.6. Visualization

In this paper we have used Matlab for visualization. We first load in the crystal cell occupation array on the fine grid, complete a 24-fold duplication to the symmetry positions which were not part of the fine grid, and convert to collections of $(x, y, z)$ triplets. We then draw partially transparent faces of a hexagonal prism around each occupied position using the Matlab patch command, omitting faces that abut other filled hexagonal prisms, and we emphasize visible edges with dark lines. If we visualize a cluster of such hexagonal prisms, as in Fig. 1*a*, the prism facets appear jagged at small scales. This is a visually irrelevant artifact of the cell shape that we avoid by altering the prism faces of horizontally-adjacent filled cells slightly so that they produce a flat prism facet. We typically display the crystal from a perspective 50 degrees above the horizontal plane to give top-down view with some three-dimensional feeling, and often supplement with a side view from within the horizontal plane and perpendicular to the widest arm diameter to show the height profile.

In the images the darkest lines are directly visible edges, while the lighter lines are visible through the crystal either as internal structure of from the back side. Fig. 4 shows a crystal with detailed interior structure shown using light grey edge emphasis, as well as some directly-visible surface and side structure shown with dark black edge emphasis. The thickness of the individual layer boundaries indicates the cell size. While drawing lines on the edges shows the surface details very well, it also gives them too much visual prominence compared to photographs. For crystals with a lot of surface structure this process will tend to make the surface too black, obscuring all detail. We have experimented with using POV ray which does not suffer this drawback, instead using cast shadows or transmitted light to depict three-dimensional information. In the end found that the Matlab images were more directly informative of the structure, although they should not be taken too visually literally.

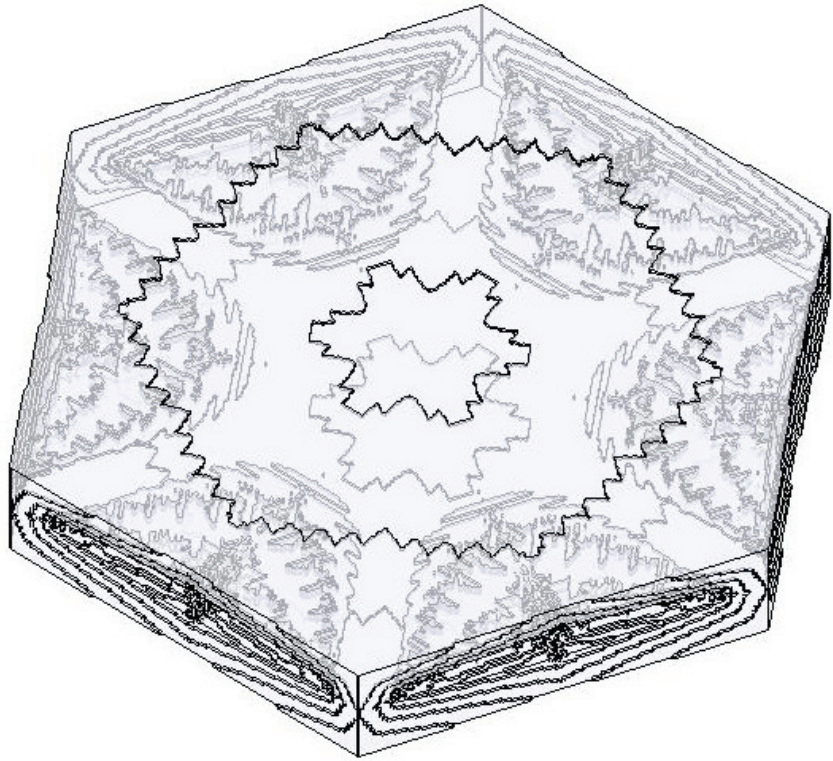

**Fig. 4. A simulated snow crystal illustrating the visual appearance of directly visible edges, shown dark, and interior edges visible through a the top surface, shown grey. The same crystal appears as part of Fig. 10.**

### 4.7. Physical Constants

The values of the physical constants used in our simulations are $c_{sat} = 10^{-6} c_{solid}$, $v_{kin} = 133$ $\mu$m/s, $\sqrt{kT/2\pi m} = 133$ m/s, and $D = 2 \times 10^{-5}$ m$^2$/s. All of these were obtained or derived from [16].

## 5. Results

We first study the crystal morphologies that result from using constant condensation coefficients following the results obtained in section 3.5. We vary the values of the condensation coefficients and study the effect on crystal morphology. We also study the effect of varying the far boundary supersaturation level and the grid size $\Delta x$. Next we study the effect of including condensation coefficients that vary with supersaturation as in nucleation-limited growth, following the results of section 3.6.

## 5.1. Constant condensation coefficient studies

**Morphological insensitivity to ambient supersaturation $\sigma_\infty$**

One remarkable aspect of the constant condensation coefficient simulations is the complete independence of the crystal morphology from the far boundary ambient supersaturation. This is illustrated in Fig. 5, which displays a crystal grown on a $\Delta x = 1 \mu m$ grid to a radius of approximately $140 \mu m$ with all $\alpha_{hv} = 1$ except $\alpha_{10} = \alpha_{20} = \alpha_{11} = 0.7$ and $\alpha_{01} = 0.1$. The far boundary ambient supersaturation level during the simulation is increased from left to right by factors of 10: $\sigma_\infty = 10^{-3}, 10^{-2}, 10^{-1}$, and $10^0$. The condensation coefficient values were chosen to result in interesting surface detail, but the more remarkable feature is that the resulting morphologies are indistinguishable.

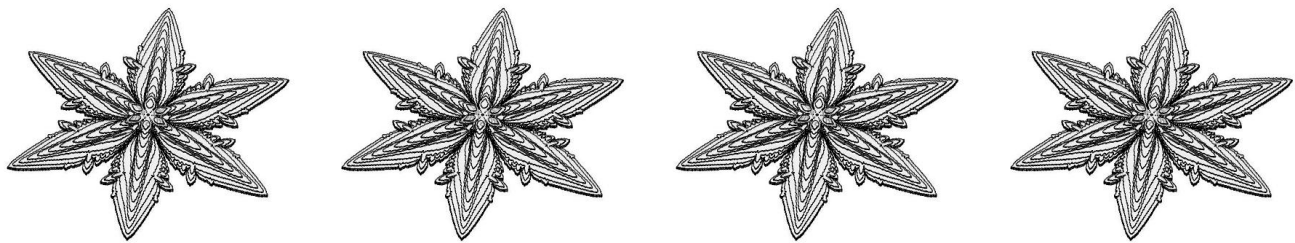

**Fig. 5. Simulated snow crystals grown with identical constant condensation coefficients at four different ambient supersaturation levels, from left to right $\sigma_\infty = 10^{-3}, 10^{-2}, 10^{-1}, 1$.**

The independence from the far supersaturation results from the combined linearity in $\sigma$ of Laplace's equation and the vapor removal condition at the crystal boundary. If $\sigma$ is doubled at the far boundary then it doubles everywhere and the same morphology results, except that the crystal grows more quickly in real time. Clearly this is not in accord with the experimental evidence, summarized in the snow crystal morphology diagram [12], showing a strong morphological dependence on the ambient supersaturation. It suggests that models containing condensation coefficients that are independent of the vapor density are limited in their explanatory power. Despite this deficiency we continue to explore the constant condensation coefficient cases in order to see what kinds of morphological features appear, and without including $\sigma_\infty$ in parameter space exploration.

**Relative insensitivity to grid size $\Delta x$**

Since the LCA grid does not represent anything physical we should hope that the morphologies end up being more-or-less insensitive to the size of the grid cell Δ*x*, and this does turn out to be the case. It would be nice to illustrate a crystal grown out to the same physical size on grids with a large range of cell sizes under otherwise equivalent simulation conditions. Unfortunately it is only computationally feasible to grow a crystal to a large physical size on a grid with a large cell size. We strike a compromise in Fig. 6, which displays a crystal grown out to various extents on various grid sizes. On the horizontal axis the cell size Δ*x* used in the simulation is doubled successively from 1 micron to 16 microns, while on the vertical axis the physical radius *R* to which the crystal is grown out is doubled from 138 microns to 1104 microns. Pairs of crystals on the same horizontal line represent the same size crystal grown on different cell size grids. For a fair comparison the initial crystal cell configuration was altered to be an approximate hexagonal prism of the same physical extent on both grids, giving 95 initial crystal cells on the smaller cell grid compared to 21 cells on larger cell grid. Pairs of crystals on the same vertical line represent crystals grown on the same cell size grid, but using a slightly different-sized initial crystal cell hexagonal prism configuration and grown to twice the horizontal extent in the upper image. The image magnification goes up by a factor of two as one moves down vertically.

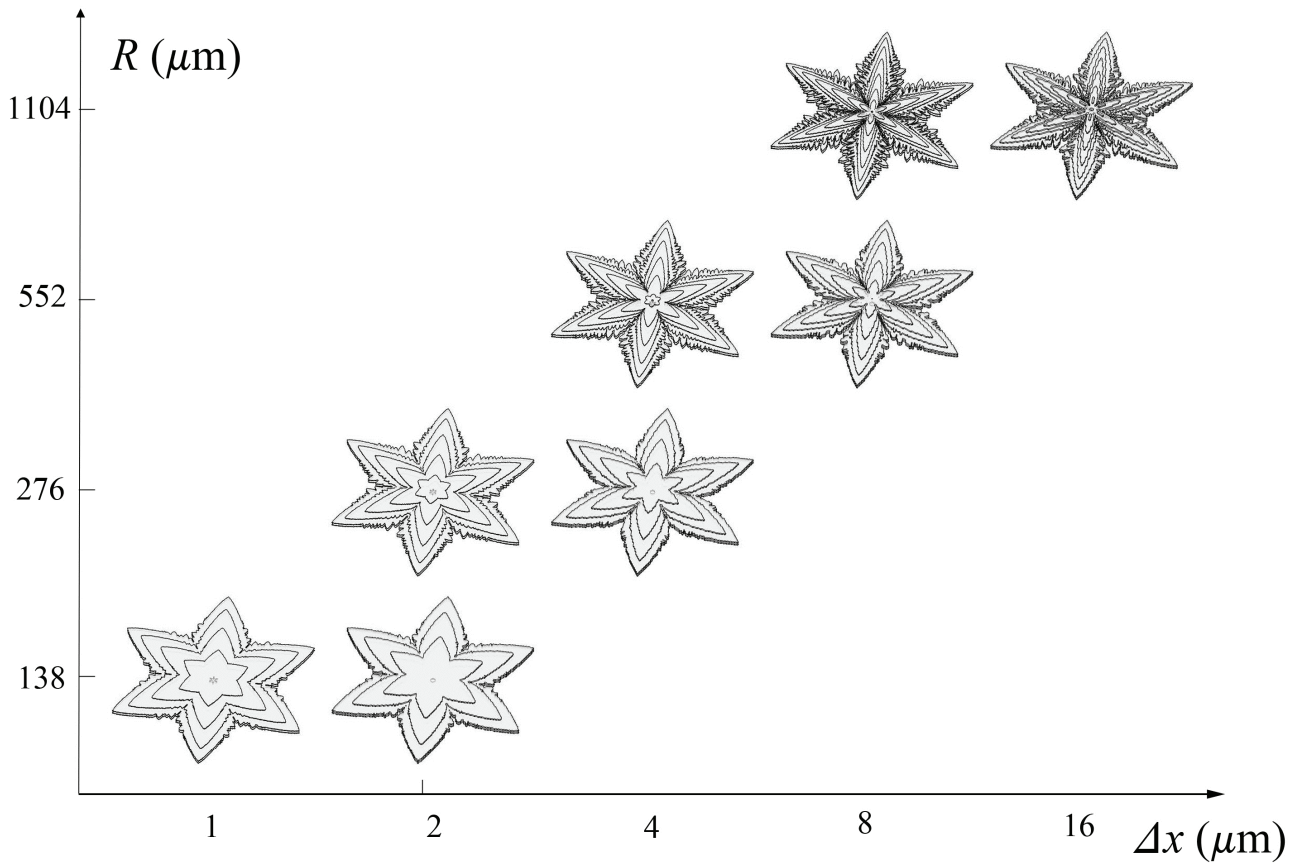

**Fig. 6. A single crystal grown with constant condensation coefficients to various horizontal radii *R* (vertical axis) on grids with varying cell sizes Δ*x* (horizontal axis).**

If the dynamics were completely independent of grid cell size then horizontally separated crystals should appear identical, except that one would be slightly more coarse grained. There are some visual differences, but overall we find the agreement to be surprisingly good. The condensation coefficients used for all of the crystals shown are $\alpha_{HV} = 1$ except for $\alpha_{01} = 0.01$, which suppresses vertical growth. These coefficients were not specially chosen except for simplicity and for the development of ridges on the arms at larger sizes.

**Morphological dependence on condensation coefficients**

We now investigate the sensitivity of the final crystal morphology to the condensation coefficients $\alpha_{HV}$ values for the case where all $\alpha_{HV}$ are independent of $\sigma$. In order to make a reasonable parameter space search we limit our figures to the reduced, two-dimensional parameter space of $\alpha_{01}$ (controlling the vertical growth rate) versus $\alpha_{10} = \alpha_{20} = \alpha_{11}$ (controlling the horizontal growth rate), with all other condensation coefficients set equal to 1 due to large numbers of neighbors.

In Fig. 7 we display crystal morphologies in this parameter space using logarithmic axes. Each larger crystal image is shown from an oblique angle of 50 degrees above the horizontal plane. To its upper right and at a smaller scale the same crystal is shown viewed from in the horizontal plane perpendicular to the maximum diameter direction in order to better indicate the height profile. To its upper left is the rough real growth time in seconds assuming ambient supersaturation of $\sigma_\infty = 0.2$. The simulations use a fine grid array size of 141x72x22 cells of size $\Delta x = 1\mu m$. Each simulation is terminated when the crystal reaches a horizontal radius of 138 cells or a vertical half-height of 20 cells. This means that the thin, needle-like growth forms reached their limits rather soon, as the grid was chosen to accommodate flat plate growth.

The overall structure of the figure confirms that the relative sizes of $\alpha_{10} = \alpha_{20} = \alpha_{11}$ and $\alpha_{01}$ determine the general preference for horizontal or vertical growth habits. $\alpha_{01}$ must be relatively suppressed by about a factor of ten to generate a more horizontal habit. For large $\alpha_{10} = \alpha_{20} = \alpha_{11}$ we find that larger $\alpha_{01}$ values generate more surface structure including arm ridges, as well as a more dendritic appearance. One can also perceive the thinning of the ridges as $\alpha_{01}$ increases, and the development of new vertical growth layers from the center which spreads outward along the arms toward the edge of the crystal. For mid-range $\alpha_{10} = \alpha_{20} = \alpha_{11}$ values we see the tendency to develop interior cavity structures on what would have been the prism facets that become deeper as $\alpha_{01}$ increases. We also see that new vertical growth layers now begin at the far points on the crystal. A further reduction in $\alpha_{10} = \alpha_{20} = \alpha_{11}$ results in faceted hexagonal prisms.

For comparison we display a similar parameter space survey in Fig. 8 using the same number of cells but instead $\Delta x = 16\ \mu m$. Although not explicitly labeled, the size scale of these crystals is a factor of sixteen larger than the scale of Fig. 7, so that the widest crystals have a radius of approximately 2.2 mm. If one believes that the morphology is not strongly affected by cell size then these crystals can be interpreted as being grown out to sixteen times the size of those in Fig. 7, excepting that the physical extent of the initial hexagonal prism configuration was sixteen times larger. At least three trends are apparent. First, there is an increased tendency toward vertical growth as growth progresses, with more of the crystals hitting the vertical fine cell boundary before the horizontal one. Second, the vertical growth develops from a central protrusion which has a size that seems to scale with the arm ridge width. Third, several of the morphological types that are present at smaller scales in Fig. 7 develop for different parameter values at larger scales in Fig. 8.

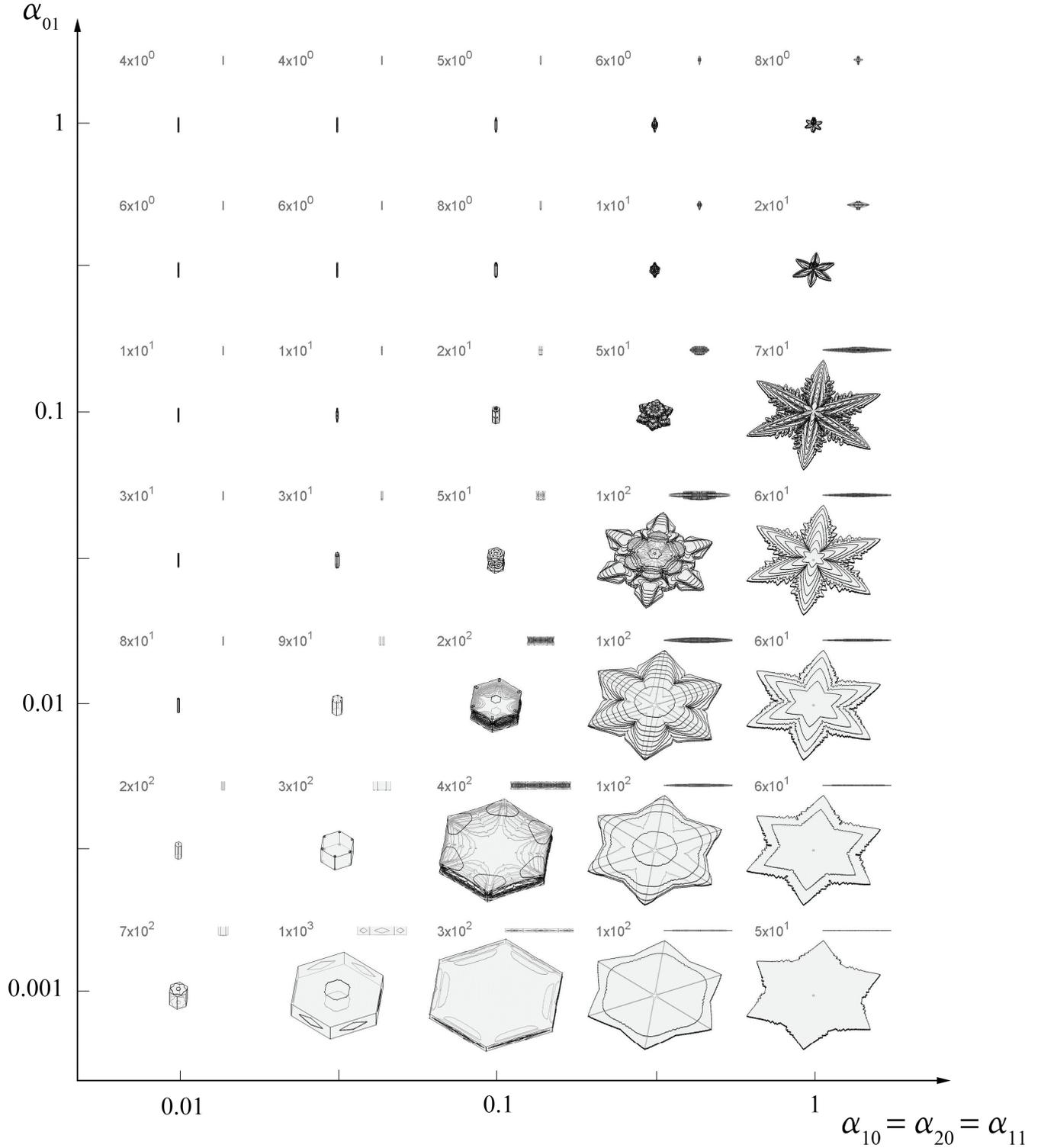

**Fig. 7. Parameter space sampling for constant condensation coefficient attachment on a $\Delta x = 1\mu m$ grid of radius 0.14 mm. All $\alpha$ values not displayed are equal to 1.**

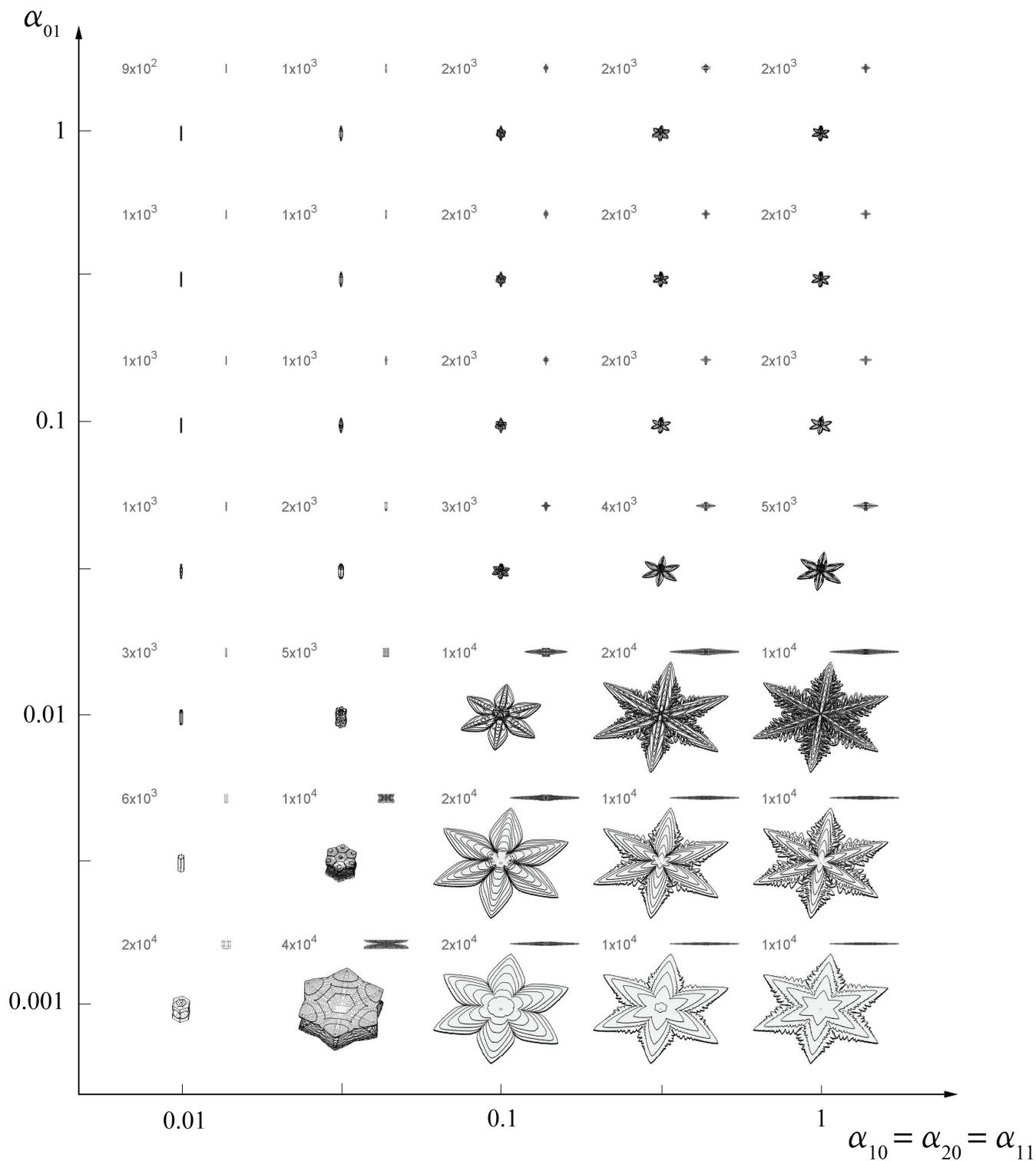

**Fig. 8. Parameter space sampling for constant condensation coefficient attachment on a $\Delta x = 16\ \mu$m grid of radius 2.2 mm. All $\alpha$ values not displayed are equal to 1.**

## 5.2. Studies with some nucleation-limited condensation coefficients

We now turn to a case where some of the $\alpha_{HV}$ are allowed to depend on the local supersaturation, and hence on $\sigma_\infty$. We use the nucleation-limited piecewise-linear model for basal facet growth introduced in section 3.6, in which $\alpha_{01}$ is linear in $\sigma$ until a value $\sigma_{01} = 0.03$, at which $\alpha_{01}$ reaches 1 and stays at 1 for larger $\sigma$ values. We use a similar model very loosely corresponding to nucleation-limited prism growth with $\alpha_{10} = \alpha_{20}$ variation determined by a free parameter $\sigma_{10} = \sigma_{20}$. The reason for setting these two condensation coefficients equal is partly aesthetic: with a constant $\alpha_{10}$ we found that tip growth became excessively suppressed relative to the arm when $\sigma_\infty$ became small, resulting in a dimpled tip of the arm. We are aware of no reason to assume that a constant condensation coefficient is a good model for tip growth, so in the absence of any alternative model we set the two equal to avoid drastic visual effects. All other condensation coefficients were held constant at 1 due to larger numbers of neighbors.

In Fig. 9 we display a plot of crystal morphology versus $\sigma_\infty$ and $\sigma_{10} = \sigma_{20}$ for simulations on a grid of 141x72x22 cells of size $\Delta x = 1$ $\mu$m. The simulation stopping criteria as well as the viewing perspectives and real-time growth in seconds are the same as in the previous subsection. Here we see strong morphological dependence on $\sigma_\infty$, with a preponderance of vertical growth for large $\sigma_\infty$, which pushes $\alpha_{01}$ to 1, and for large $\sigma_{10} = \sigma_{20}$, which suppresses horizontal growth. Arm ridges are present at high $\sigma_\infty$ and narrow as $\sigma_\infty$ increases. At higher $\sigma_{10} = \sigma_{20}$ as $\sigma_\infty$ is decreased the six arms split at the central plane into twelve as upward growth transitions to the arm tips, and with decreasing $\sigma_\infty$ the arms fuse while retaining the internal structure, and finally become hexagonal prisms at very low $\sigma_\infty$. For intermediate $\sigma_{10} = \sigma_{20}$ values the arm splitting does not occur but the interior structure develops, while at low $\sigma_{10} = \sigma_{20}$ no interior structure develops as $\sigma_\infty$ decreases. Overall the model seems to resist facetization, requiring very small $\sigma_\infty$ before unblemished facets occur, and with correspondingly outlandish growth times.

Again for comparison we display a similar parameter space survey in Fig. 10 using the same number of cells but instead $\Delta x = 16$ $\mu$m. Like the comparison in section 5.1 the size scales here are 16 times larger than in Fig. 9, and might be loosely interpreted as the same crystals grown out to sixteen times the diameter, but with slightly different initial conditions. Here we see that some of the crystals at larger $\sigma_\infty$ that appeared to have a stable horizontal growth habit in Fig. 9 later tend toward more vertical growth. Also, again we see that morphologies that are present at smaller scales in Fig. 9 tend to develop at larger scales for different parameter values in Fig. 10. Finally, the very low $\sigma_{10} = \sigma_{20}$ crystal that were growing as hexagonal prism at $\Delta x = 1$ $\mu$m have undergone instabilities creating interior cavities or arms at larger scales.

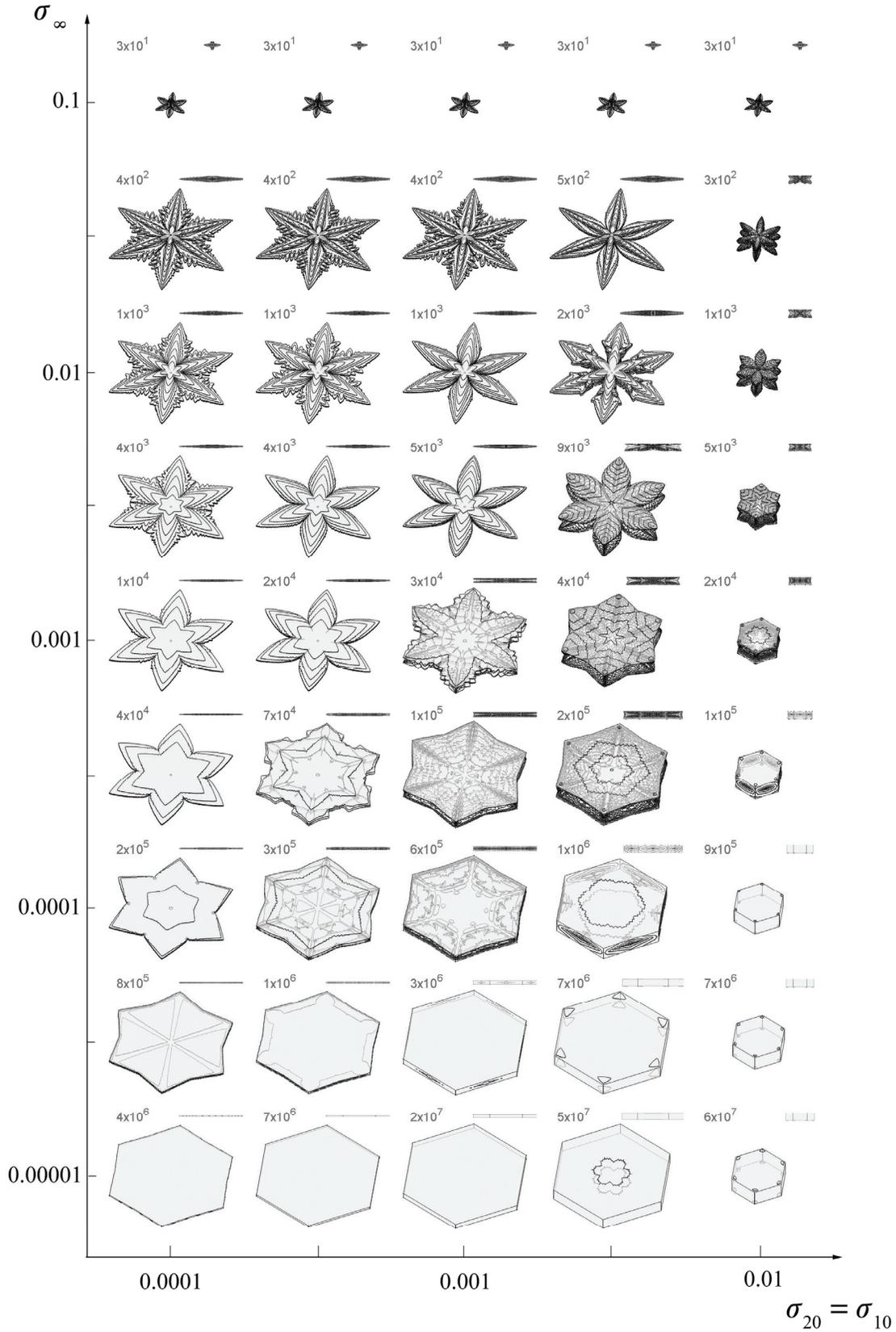

**Fig. 9. Parameter space sampling for nucleation-limited condensation coefficient model on a $\Delta x = 1$ $\mu$m grid of radius 0.14 mm.**

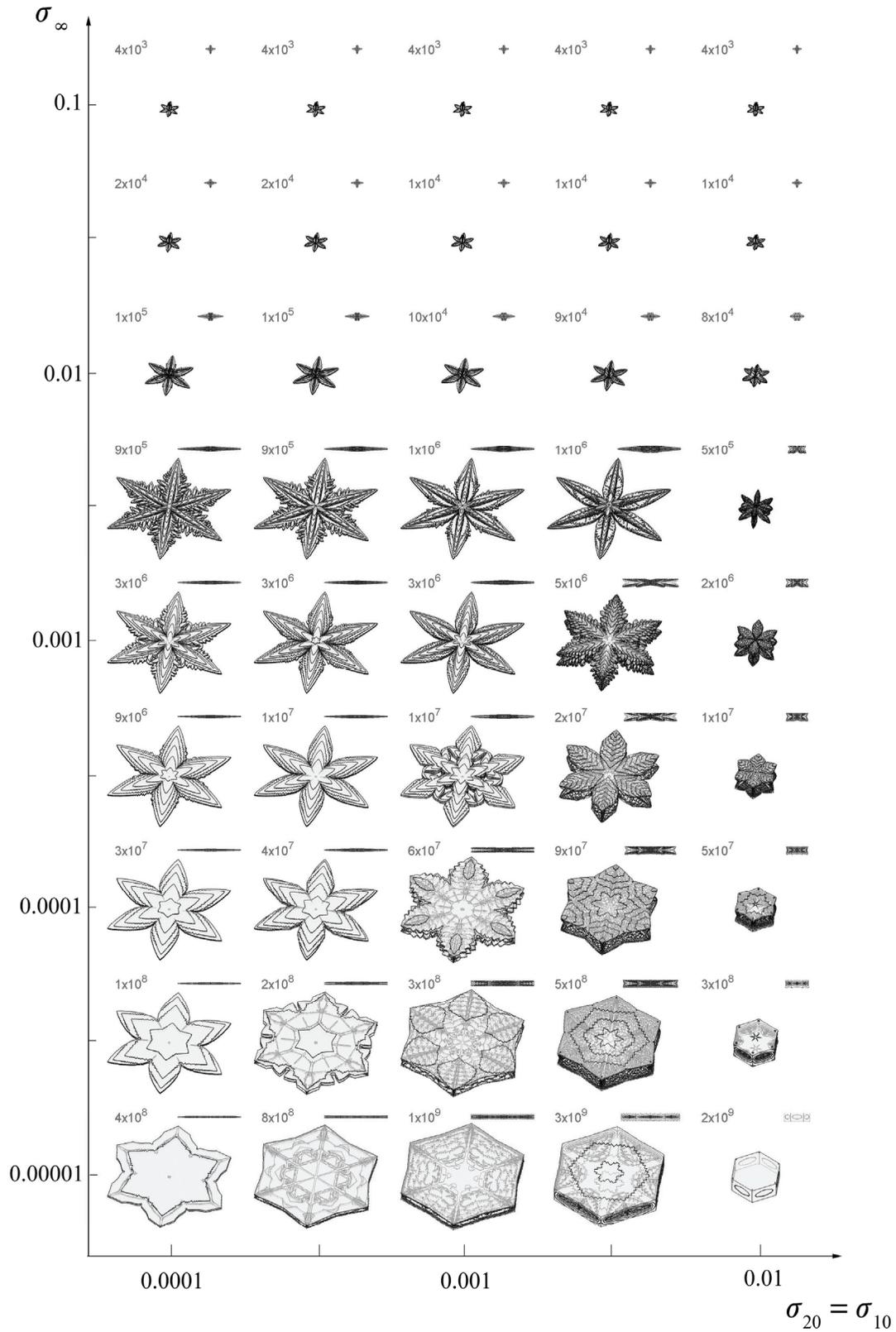

**Fig. 10. Parameter space sampling for nucleation-limited condensation coefficient model on a $\Delta x = 16\ \mu$m grid of radius 2.2 mm.**

# 5. Conclusions

Our goal in this work was to make changes to the Gravner-Griffeath model that move the simulation in the direction of increased physical realism, while maintaining a computationally feasible fully three-dimensional model of snow crystal growth. We now refer back to the weaknesses identified at the end of section 2 to reflect on the degree to which our model has been successful, and to identify remaining weak points.

We have tied the diffusion and attachment dynamics to an explicitly-defined grid cell size, guided by the discussion and results in [16]. The problem of excessive supersaturation draining has been avoided by decoupling the growth and vapor diffusion, and by keeping the vapor distribution in equilibrium via relaxation. The growth rate of the crystal is now based on the Hertz-Knudson formula. This forms one of the remaining physically weak aspects of our model. The formula was intended to apply to rough surfaces, or to flat facets when the nucleation-limited modification is used, yet here we apply it to cases involving facet edges and corners. Indeed, the algorithm is sensitive only to neighborhood properties, and thus can only make a poor guess as to whether the crystal is faceted or rough. Still, the correct connection with some growth domains is an improvement. The vapor removal rate at the crystal surface has been tied to the growth rate by mass conservation. The quasiliquid layer used in the Gravner-Griffeath model has been eliminated, although we provide no evidence that our model is equally expressive without it. The potential problem of high initial vapor density affecting morphology has been eliminated by keeping the vapor distribution in equilibrium at every stage of growth. The problem of too many free parameters remains in our model since the condensation coefficients depend on the local crystal geometry, much as in the Gravner-Griffeath model.

The figures in the results section exhibit several interesting and promising features. The independence of growth morphology from $\sigma_\infty$ in the constant condensation coefficient case was surprising to us but is easily understood. The relative insensitivity of morphology to grid cell size is reassuring and would seem to be a necessary property for any model implementing dynamics at a mesoscopic scale. The characteristic arm ridges and interior structures produced by the Gravner-Griffeath model persist in our model morphologies. The ridge width is inversely correlated with the tendency for vertical growth, and there is some indication that the presence of ridges indicates a long-term instability with respect to vertical growth, although it may never be realized under real growth conditions. The interior structures appear more varied and detailed than the surface structures, which are more or less limited to ridges on dendritic structures. We suspect that rapidly spatial variation of the supersaturation near a cavity opening may encourage more dramatic growth instabilities which result in these structures. Finally, a scaling relation between parameter choices and grid size seems to be apparent in both our constant $\alpha$ and varying $\alpha$ studies.

At the same time some possible model weaknesses are hinted at by the resulting morphologies. One is the difficulty with which facetization is achieved by our model. Model parameters must be made quite small in order to produce facets, and the growth times can become correspondingly outlandish. Our model also seems to fail to produce ridges in combination with facetization, ridges on plate-like structures, and faceted dendrites. These combinations are clearly observed in snow crystal photographs and also appear in the Gravner-Griffeath model morphologies. Comparison with natural snow photographs can be somewhat misleading because the varied morphologies observed are partly due to the range of growth conditions encountered during the growth period, including changes in temperature and supersaturation levels. It is possible that the facetization observed on partly dendritic natural crystals could be the result of a falloff of ambient vapor density toward the end of the growth period. We did not attempt to reproduce this effect in our simulations.

## References

[1] K. G. Libbrecht, "Ken Libbrecht's Field Guide to Snowflakes", Voyageur Press, 2006.

[2] K. G. Libbrecht, P. Rasmussen, "The Snowflake: Winter's Secret Beauty", Voyageur Press, 2003.

[3] K. G. Libbrecht, `http://www.its.caltech.edu/~atomic/snowcrystals/`

[4] W. A. Bentley, W. J. Humphreys, "Snow crystals," McGraw-Hill Book Company, 1931. Also, New York: Dover Publications; 1962.

[5] U. Nakaya, "Snow Crystals: Natural and Artificial", Harvard University Press, 1954.

[6] T. Witten, L. Sander, *Diffusion-limited aggregation, a kinetic critical phenomenon*, Phys. Rev. Lett. 47 (1981), 1400-1403.

[7] N. H. Packard, *Lattice models for solidification and aggregation*, in "Theory and Application of Cellular Automata," World Scientific, 1986, pp. 305-310.

[8] R. Xiao, J. Alexander, F. Rosenberger, *Morpological evolution of growing snow crystals: a Monte Carlo simulation*, Phys. Rev. A 38 (1988), 2447-2456.

[9] R. Kobayashi, *Modeling and numerical simulations of dendritic crystal growth*, Phys. D. 63 (1993) 410-423.

[10] J.-M. Debierre, A. Karma, F. Celestini, R. Guérin, *Phase field approach for faceted solidification*, Phys. Rev. E 68, (2003) 041604-1-13.

[11] J. W. Barrett, H. Garcke, R Nürnberg, *Numerical computations of facetted pattern formation in snow crystal growth*, Phys. Rev. E **86** (2012), 011604.

[12] K. G. Libbrecht, *The physics of snow crystals*, Rep. Prog. Phys. **68** (2005) 855-895.

[13] C. A. Reiter, *A local cellular model for snow crystal growth*, Chaos Solit. Fract. 23 (2005) 1111-1119.

[14] J. Gravner, D. Griffeath, *Modeling snow crystal growth: a three-dimensional mesoscopic approach*, Phys. Rev. E **79** (2009) 011601.

[15]  H. R. Pruppacher and J. D. Klett, "Microphysics of Clouds and Precipitation," Kluwer Academic Publishers, 1997.

[16]  K. G. Libbrecht, *Physically Derived Rules for Simulating Faceted Crystal Growth using Cellular Automata*, arXiv:0807.2616 (2008).

[17]  Y. Saito, "Statistical Physics of Crystal Growth", World Scientific, Singapore, 1996.

## Acknowledgements

We thank Ken Libbrecht and David Griffeath for clarifications of their models, Chris Robinson for his early work on the simulation code, and John Kieffer for his work on visualization.